\documentclass[doublecol]{epl2} 
\usepackage{epsfig}
\usepackage{amsmath}
\usepackage{amssymb}

\usepackage[pdftitle={KierfeldDraft09-11-13.pdf},
            pdfauthor={Jan Kierfeld},
        pdfkeywords={Coupling of actin hydrolysis and 
        polymerization in a two-state model}
            ]{hyperref}

\title{Coupling of actin hydrolysis and polymerization: Reduced description
   with two nucleotide states}
\shorttitle{Coupling of actin hydrolysis and polymerization} 
\author{Xin Li\inst{1,2}, Reinhard Lipowsky\inst{1} \and   Jan Kierfeld\inst{3}
}
\shortauthor{J. Kierfeld  \etal}

\institute{
 \inst{1} Max Planck Institute of Colloids and Interfaces, Science
  Park Golm, 14424 Potsdam, Germany, EU\\
  \inst{2} Key Laboratory of Frontiers in Theoretical Physics, ITP,  CAS, 
    Beijing 100090, China\\
  \inst{3} Physics Department,
  TU Dortmund University, 44221 Dortmund, Germany, EU
}

\pacs{87.16.Ka}{Filaments, microtubules, their networks, and 
         supramolecular assemblies}
\pacs{87.16.A-}{Theory, modeling, and simulations}
\pacs{87.15.rp}{Polymerization}

\abstract{
The polymerization of actin filaments 
is coupled to the hydrolysis of adenosine triphosphate (ATP),
which involves both the cleavage of ATP and the 
release of inorganic phosphate. 
We describe hydrolysis by a reduced  two-state model
with a  cooperative cleavage mechanism, where the cleavage rate 
 depends on the   
state of the neighboring actin protomer in a filament. 
We obtain theoretical predictions of
experimentally accessible steady state quantities such as the 
size of the ATP-actin cap,  the size distribution of ATP-actin 
islands, and the cleavage flux for  
cooperative cleavage mechanisms.
 }

\begin{document}

\maketitle

%%%%%%%%%%%%%%%%%%%%%%%%%%%%
\section{Introduction}
%%%%%%%%%%%%%%%%%%%%%%%%%%%%

Actin filaments are an important structural element of the 
cytoskeleton, and  their ATP-driven polymerization dynamics 
plays an important role in  cell motility \cite{bray}. 
The  ATP hydrolysis in actin filaments  is the 
basis for treadmilling, i.e., 
the  simultaneous polymerization and 
depolymerization at the two ends of a filament,  and is necessary 
for actin-mediated force generation and motility \cite{theriot00}.

Actin monomers (G-actin) assemble into polar actin filaments 
(F-actin) with a  fast polymerization dynamics at the 
  barbed end and 
a  slow  polymerization dynamics at the  pointed end. 
Actin monomers can bind ATP, which is then  
hydrolyzed in a two-step process into adenosine diphosphate (ADP) 
and inorganic phosphate ($\rm P_i$). First, 
 ATP is cleaved into the complex ADP-$\rm P_i$, from which 
$\rm P_i$ is released in a second step.
After incorporation into a filament the actin protomers can therefore be 
in three different states: a T-state (ATP-actin), a $\Theta$-state 
(ADP-$\rm P_i$-actin), and a D-state (ADP-actin).

Past studies have focused 
on two sorts of cleavage processes in actin filaments. 
In {\em random cleavage}, 
T-protomers are cleaved independent of the state of 
their neighbors \cite{blanchoin02,VYOS05,fuji07}.
In {\em  vectorial cleavage}, there is a sharp 
interface between the ATP-cap containing only T-protomers 
and the remaining filament consisting of $\Theta$- and D-protomers,
and cleavage can only occur at the T$\Theta$-interface
 \cite{PCK85,korn87,melki96,SK06}.

In this Letter, we investigate {\em cooperative} hydrolysis mechanisms, 
where the cleavage  rate of each monomer 
depends on the state of its neighbors and which contain
 random and vectorial mechanisms as special cases. 
Cooperative hydrolysis was previously discussed in 
Ref.\ \cite{carlier87} for actin and Ref.\ \cite{flyvbjerg94} for 
microtubules.
One important piece of evidence for a cooperative hydrolysis is the
small hydrolysis rate of G-actin as compared to 
F-actin, which 
 acts as a very effective  ATPase with a fast hydrolysis rate. 
This pronounced change of the ATP hydrolysis rate after inclusion of 
G-actin monomers into filaments  suggests that the cleavage 
rate  is affected by binding to other protomers 
in the filament. 
Furthermore, 
structural differences between  T- and D-state protomers 
 indicate that cleavage might be affected 
by the same structural elements that are also involved in the binding 
of protomers \cite{graceffa03,rould06}.

A complete model of ATP hydrolysis involves all three nucleotide 
states of actin protomers. 
We have studied such three-state models for cooperative 
ATP hydrolysis  in Ref.\ \cite{XLJKRL}.
In this Letter, we will consider a reduced two-state model
by combining the $\Theta$-state  and the D-state of protomers 
 into a single ${\rm D}^*$-state.
This means that we  
focus on the cleavage 
of T-protomers and  ignore the subsequent 
process of $\rm P_i$-release.
The reduction to two protomer states can 
 be justified for fast growth 
at high T-monomer concentrations. 
The reduced  model has the advantage that we are able 
to obtain analytic predictions for important steady state
observables, such as the 
size of the ATP-actin cap, the size distribution of ATP-actin 
islands, and the cleavage flux.
In particular, we can calculate analytically how these quantities
depend on the  
cooperativity of the cleavage mechanism.  
We find an intriguing scaling behavior in the limit of 
strongly cooperative cleavage.

%%%%%%%%%%%%%%%%%%%%%%%%%%%%%%%%%%%%%%%%%
\section{Two-state model for cooperative hydrolysis}
%%%%%%%%%%%%%%%%%%%%%%%%%%%%%%%%%%%%%%%%%

\begin{figure}
  \begin{center}
  \epsfig{file=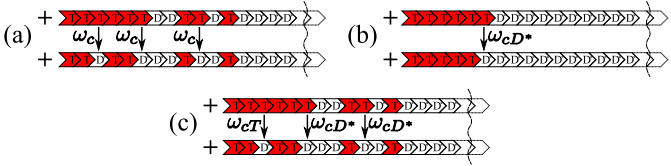,width=0.49\textwidth}
  \caption{
 Actin filament consisting of  T-protomers (red) and 
$\rm D^*$-protomers (white) in the reduced two-state model. 
(a) Random cleavage with rate $\omega_c$; (b) vectorial cleavage
with rate $\omega_{c,D^*}$; 
 (c) cooperative cleavage with rate $\omega_{c,D^*}$ and $\omega_{c,T}$
depending on the local neighborhood.
    }
  \label{model}
  \end{center}
\end{figure}

In the reduced two-state model, we combine
 the $\Theta$-state  and the D-state of protomers 
 into a single ${\rm D}^*$-state and
model the actin filament as a one-dimensional sequence 
of actin monomers in the T- or ${\rm D}^*$-state (ignoring the helical 
structure of the filament), see Fig.~\ref{model}.

In general, 
 cleavage within the actin filament happens according to a 
cooperative mechanism, i.e., the cleavage rate of each monomer 
depends on the state of its neighbors. 
Each monomer is {\em polar}, therefore we do not  expect 
 cleavage and release 
 rates to be  mirror symmetric, i.e., they need not to be 
 invariant under exchange of the two neighbors. 
The simplest way to introduce cooperativity 
without mirror symmetry is to assume that the cleavage rate depends 
on the state of the neighbor in the direction of one of the ends. 
Because the ATP-binding cleft is located 
in the direction of the pointed end, it is plausible to assume 
that cleavage depends on the state of the neighbor monomer on the 
``pointed side''. 
Consequently, we introduce two cleavage rates for an ATP monomer: 
$\omega_{cT}$ if the monomer has a  T-neighbor on the pointed side
and $\omega_{cD^*}$ 
if the monomer has a ${\rm D}^*$-neighbor on the pointed side. 
The two cleavage rates   $\omega_{cT}$, $\omega_{cD^*}$, 
 can also be written as 
\begin{equation}
    \omega_{cD^*}\equiv \omega_c~~,~~
    \omega_{cT} \equiv \omega_c\rho_c,
\label{wa}
\end{equation}
which defines the {\em cleavage parameter} $\rho_c$. 
We assume that the presence of a cleaved  neighbor monomer 
{\em increases} the cleavage rate, which implies  $\rho_c\le 1$. 
 In the special case of $\rho_c=1$, the cleavage rate does 
{\em not} depend on the state of neighboring monomers,
which  corresponds to {\em random} cleavage.
In the limiting case of  $\rho_c=0$, cleavage {\em only} happens
at the ${\rm TD}^*$-interface within the filament corresponding to a
{\em vectorial} cleavage mechanism.
A small cleavage parameter $\rho_c \ll 1$ corresponds to a 
strongly cooperative cleavage mechanism.

In the following, we will focus on  the barbed end of the filament. 
T-monomers attach with a rate $\omega_{\rm on}$ (the number 
of T-monomers attaching per unit time) at the barbed end. 
This attachment rate 
is proportional to the concentration $C_T$ of T-monomers in solution
and the rate constant $ \kappa_{\rm on}$, i.e., 
$\omega_{\rm on}= \kappa_{\rm on}C_T$. 
We assume that the attachment of ${\rm D}^*$-monomers is not possible, 
i.e., $\omega_{\rm on,D^*}=0$, which is justified in view of smaller
rate constants \cite{fuji07} and small concentrations 
of $\Theta$- and D-monomers in solution.

Both $T$- and ${\rm D}^*$-monomers can detach from the barbed 
end with rates $\omega_{\rm off,T}$ and $\omega_{\rm off,D^*}$. 
Starting from the full three-state model,
an effective 
detachment rate $\omega_{\rm off,D^*}$  of ${\rm D}^*$-monomers
can only be defined consistently for the whole 
T-monomer concentration range $C_T$ if
$\Theta$- and D-monomers have similar detachment rates, 
$\omega_{\rm off,\Theta}\simeq \omega_{\rm off,D}$.
Measured values  $\omega_{\rm off,\Theta}\simeq 0.2s^{-1}$ and  
$\omega_{\rm off,D}\simeq 5.4s^{-1}$ \cite{fuji07} show that this 
condition is violated. 
Nevertheless, a reduced two-state model can still be introduced for 
high T-monomer concentrations $C_T$, 
where the probability $P_{1,D^*}$
 that the first protomer at the barbed end is 
a ${\rm D}^*$-protomer is negligible 
or the  probability $P_{1,T}$
 that the first protomer at the barbed end is 
a $T$-protomer is close to one, 
\begin{equation}
    P_{1,D^*}\approx 0 ~~\mbox{and}~~  P_{1,T}=1-P_{1,D^*} \approx 1.
\label{smallP1D}
\end{equation}
In this limit 
the corresponding detachment flux  of ${\rm D}^*$-monomers
$P_{1,D^*}\omega_{\rm off,D^*}\approx 0$ is always negligible, and 
 the  detachment rates of both  $\Theta$- and D-monomers 
become irrelevant for the polymerization process.
In the following we will focus on this limit $P_{1,D^*} \approx 0$. 
We will show below, see eq.\ (\ref{P1D}), that it 
  is realized for strongly cooperative 
cleavage
$\rho_c\ll 1 $ or for fast growth 
at high T-monomer concentrations $C_T$.

We do not take into account a possible cooperativity in the 
attachment and detachment process, i.e., 
 $\omega_{\rm on}$,  $\omega_{\rm off,T}$, and  $\omega_{\rm off,D^*}$
do not   depend on the state of the last monomer, which is at the
 tip before attachment or which is left behind at the tip after 
detachment.
We also neglect fracture of filaments, which has been discussed for 
hemoglobin fibers in Ref.\ \cite{wang09}.
Literature values for cleavage, attachment, and detachment 
 rates of our model are listed in table 
\ref{tableparameters}.

\begin{table}
\caption{Literature values for model  parameters.}
\label{tableparameters}
\begin{center}
\begin{tabular}{|c|c|c|c|}
\hline
 $\kappa_{\rm on}(\mu M^{-1} s^{-1})$  &   $\omega_{\rm off,T}(s^{-1})$ 
   & $\omega_{\rm on,D^*}$ 
  & $\omega_c (s^{-1})$ \\
\hline
  $11.6$ \cite{TDP86,fuji07}  &   
       $1.4$ \cite{TDP86,fuji07}   & 
    0 & 
    $0.3$ \cite{blanchoin02}\\ 
\hline
\end{tabular}
\end{center}
\end{table}

We will derive analytic results for the
length  of the ATP-actin cap,  the length distribution of ATP-actin 
islands, and the cleavage  flux.
We compare these results to  stochastic simulations of  the full
 three-state model, which were performed using the Gillespie 
algorithm as described in Ref.\ \cite{XLJKRL}. 
For sufficiently  high T-monomer concentrations $C_T$, we expect 
simulation results for the three-state model to agree 
with our analytic results  for the reduced two-state model. 
In the stochastic simulations, we will use 
a three-state model with a
random $\rm P_i$-release mechanism with  release rate 
$\omega_r = 0.003 s^{-1}$ \cite{melki96,blanchoin02}. 
Furthermore, we use $\omega_{\rm off,\Theta}=  0.2s^{-1}$ \cite{fuji07}
and 
$\omega_{\rm off,D} =  5.4s^{-1}$ \cite{fuji07} for the 
off-rates of  $\Theta$- and D-protomers, respectively.

%%%%%%%%%%%%%%%%%%%%%%%
\section{Growth rate and critical concentrations}
%%%%%%%%%%%%%%%%%%%%%%%

Attachment and detachment parameters for T-monomers 
lead to the 
 T-protomer growth rate, 
\begin{equation}
 J_T = \omega_{\rm on}-P_{1,T}\omega_{\rm off,T}\approx 
   \kappa_{\rm on} C_T- \omega_{\rm off,T}.
\label{JT}
\end{equation}
Because there is no ${\rm D}^*$-attachment, $\omega_{\rm on,D^*}= 0$, 
and we  consider the  limit $P_{1,D^*}\approx 0$, 
the T-protomer growth rate  $J_T$ equals the total growth rate $J_g$ 
of the filament,
\begin{equation}
  J_{g}= \omega_{\rm on}-P_{1,T}\omega_{\rm off,T} - P_{1,D^*}\omega_{\rm
    off,D^*} \approx J_T.
\label{Jg} 
\end{equation}
The critical concentration for filament growth
at the barbed end is given by $C_{T,g}= \omega_{\rm off,T}/ \kappa_{\rm on}$
with $C_{T,g} \simeq 0.12{\rm \mu M}$ for the values given in table 
\ref{tableparameters}.

For  vectorial cleavage with $\rho_c=0$, there is a single ATP-island 
at the filament tip and a single ${\rm TD}^*$-interface, where
cleavage  takes place with a rate $\omega_{cD^*}=\omega_c$.  
The length of this ATP-tip becomes infinite in the steady state 
if the  T-protomers  growth rate exceeds this cleavage rate, 
$J_T>\omega_c$, which 
defines a threshold concentration 
$C_{T,c} = (\omega_c+\omega_{\rm off,T})/\kappa_{\rm on}$
with $C_{T,c} \simeq 0.15{\rm \mu M}$ for the values given in table 
\ref{tableparameters}.
The dimensionless growth parameter 
$J_T/\omega_c = (C_T-C_{T,g})/(C_{T,c}-C_{T,g})$
 characterizes the competition of growth and hydrolysis
currents.
Because of condition (\ref{smallP1D}),
we will  focus  on fast growth 
with $J_T/\omega_c \gg 1$ in the following,  see eq.\ (\ref{P1D}) below, 
which is realized  for T-monomer concentrations 
much larger than the corresponding 
threshold concentration, $C_T \gg C_{T,c}$.

%%%%%%%%%%%%%%%%%%%%%%%%%%%%%%%%%%%%%%%%%
\section{Length distribution of ATP-tip}
%%%%%%%%%%%%%%%%%%%%%%%%%%%%%%%%%%%%%%%%%

The  growing barbed end consists of a sequence of T- and 
${\rm D}^*$-monomers. The state of the filament can be described 
as a sequence of  connected islands of T-monomers which are separated 
by ${\rm D}^*$-monomers. Similar to  an analysis of the case of 
 random hydrolysis in microtubules in 
Ref.\ \cite{A07}  we will focus on the length distribution 
 of these ``ATP-islands''.  First we will consider the 
length of the {\em first} ATP-island at the barbed end, which we call 
``ATP-tip'' in the following.

\begin{figure}
  \begin{center}
  \epsfig{file=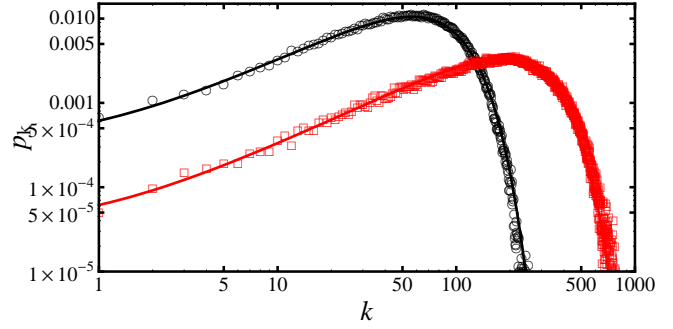,width=0.48\textwidth}
  \caption{
Length distribution $p_k$ of ATP-tip for  an actin concentration 
$C_T=1\mu M$ (corresponding to $J_T/\omega_c = 34$)
and  cleavage parameters $\rho_c=10^{-2}$ (black, $\circ$) and $10^{-3}$ 
(red, $\square$).
Other parameter values as in table \ref{tableparameters}.
Comparison between (i) analytic results from  eq. (\ref{NkAi})  
using $p_k=P_k-P_{k+1}$ (solid lines) and 
(ii)  results from  stochastic simulations 
using the Gillespie algorithm (circles).
  }
  \label{nksol}
  \end{center}
\end{figure}

The probability $p_k$ 
of finding an ATP-tip of length $k$ 
$k=0,1,2,...$, 
with $k=0$ corresponding to the case of a  $\rm D^*$-monomer right 
 at the barbed end,
satisfies the master equation 
\begin{eqnarray}
  \partial_t p_k &=& J_T (p_{k-1} - p_k) 
    -\omega_c \left[ 1 + \rho_c(k-1) \right]p_k  (1-\delta_{k,0})  
   \nonumber\\
   &&
    + \omega_c p_{k+1}+ \omega_c\rho_c \sum_{s \ge k+2}p_s
\label{nkcoop}
\end{eqnarray}
with boundary condition $p_{-1} \equiv 0$.
The first term on  the rhs of eq.\ (\ref{nkcoop}) 
describes loss and gain by 
attachment of T-protomers. 
The second term is nonzero only for tip lengths 
$k>0$ and  describes  loss by cleavage: the last T-monomer 
at the  ${\rm TD}^*$-interface is cleaved 
with a rate $\omega_{cD^*}$, whereas 
the remaining  $k-1$  T-monomers of the tip 
are cleaved with a rate $\omega_{cT}$. 
The last two terms are the corresponding 
gain terms from cleavage: an ATP-tip length of $k$ can be obtained 
by cleavage at the  ${\rm TD}^*$-interface 
of an ATP-tip of length $k+1$ with rate
$\omega_{cD^*}$ , or it can be
obtained by ``fragmentation'' of a tip of length $s\ge k+2$ into two pieces 
by cleavage 
of its interior T-monomers with rate $\omega_{cT}$. 
For the special case of random cleavage with $\rho_c=1$ we recover the
results  of Ref. \cite{A07};
for vectorial cleavage with $\rho_c=0$, 
eq.\ (\ref{nkcoop}) reduces to a random walk in $k$-space ($k>0$)
with stepping probability $J_T$ from $k-1$ to $k$ and $\omega_c$ from 
$k+1$ to $k$. 
With a continuum approximation in the 
variable $k$, eq.\ (\ref{nkcoop}) has been obtained also in 
Ref.\ \cite{flyvbjerg94}.

The steady  state of the ATP-tip length distribution 
$p_k$ is obtained 
by setting the rhs of eq.\ (\ref{nkcoop}) equal to zero. 
It is convenient to consider the cumulative quantity 
$P_k \equiv \sum_{l\ge k}p_l$
with $P_0= 1$ and 
$P_1 = P_{1,T}$, for which we find the recursion  relation
\begin{equation}
   P_{k+1}-P_k  = 
   \frac{J_T}{\omega_c(1-\rho_c)} (P_{k}-P_{k-1})
    + \frac{\rho_c}{1-\rho_c} k P_k 
\label{Precursion}
\end{equation}
in the steady state with $k \ge 1$. 
For vectorial cleavage with $\rho_c=0$, we find an
exponentially decaying  stationary 
solution $p_k \sim (J_T/\omega_c)^k$ 
 for $J_T<\omega_c$, i.e.,  below the threshold concentration 
$C_T<C_{T,c}$. For  $J_T>\omega_c$, the ATP-tip is steadily growing and 
no stationary solution can be found.

For the general case of arbitrary cooperativity,
we apply a continuum approximation in the 
variable $k$,
\begin{align}
0 &= \partial_k^2P_k(k)   + a \partial_kP_k(k) -b kP_k 
~~~~~\mbox{with}\nonumber\\
  a&\equiv 2\frac{  1-\rho_c- {J_T}/{\omega_c}}
          {  1-\rho_c + {J_T}/{\omega_c}} 
 ~~\mbox{and}~~
 b \equiv  2\rho_c\frac{1}
          {1-\rho_c + {J_T}/{\omega_c}}. 
\label{Nkcoopcont}
\end{align}
The solution of this equation 
with boundary condition $P_0=1$  is
\begin{eqnarray}
P_k &=& \exp(-ak/2)
    \frac{{\rm Ai}(b^{-2/3}a^2/4+b^{1/3}k)}{{\rm Ai}(b^{-2/3}a^2/4)}
\label{NkAi}
\end{eqnarray}
where ${\rm Ai}(x)$ is the Airy function \cite{abra84}.
The ATP-tip length distribution is obtained as $p_k=P_k-P_{k+1}$ from 
the solution (\ref{NkAi}), see Fig.\ \ref{nksol}.
The  continuum approximation used to derive eq.\ (\ref{Nkcoopcont}) 
is justified  because neither $|a|$ nor $b$ can become large compared to
unity.

In the following we will focus again on fast growth 
with $J_T/\omega_c \gg 1$, which leads to  
$a\approx -2$ and $b\approx 2\rho_c\omega_c/J_T\ll 1$.
Using the asymptotics  of  the Airy function, 
${\rm Ai}(x) \sim  e^{-2x^{3/2}/3}$ for $x\gg 1$ \cite{abra84},
we find $P_k \approx e^{-bk^2/2|a|}$ and, consequently, 
the resulting tip length distribution 
$p_k \approx -\partial_k P_k(k)$ is  exponentially decaying for large $k$, 
\begin{equation}
 p_k     \approx    ({kb}/{|a|}) e^{-bk^2/2|a|}
   \approx  k
 ({\omega_c\rho_c}/{J_T})e^{-k^2\omega_c\rho_c/2J_T}.
\label{pkfast}
\end{equation}
An interesting observable, which may be experimentally accessible in 
experiments on
 single filaments,  is the mean tip size 
$\langle k \rangle = \sum_{k\ge 1} k p_k$. 
For fast growth we obtain from eq.\ (\ref{pkfast})
a characteristic square-root dependence on the 
cleavage parameter $\rho_c$, 
\begin{equation}
 \langle k\rangle \approx  \sqrt{\pi/2} (|a|/b)^{1/2}  
 \approx  \sqrt{\pi/2}  (J_T/\omega_c\rho_c)^{1/2},
\label{kmeanresult}
\end{equation}
which  could be used 
in experiments to determine $\rho_c$ by measuring the tip length.
Another experimentally accessible observable is 
the probability  $P_{1,D^*}$ that the first monomer is 
in the  $D^*$-state, for which we find 
\begin{equation}
  P_{1,D^*} = 1-P_1 
   \approx {\rho_c\omega_c}/{2J_T},
\label{P1D}
\end{equation}
i.e., a linear dependence on $\rho_c$.
Eq.\ (\ref{P1D}) also confirms that the  
limit  $P_{1,D^*}\approx 0$, see (\ref{smallP1D}),
 is attained for strongly cooperative cleavage 
$\rho_c\ll 1 $ or for fast growth $J_T \gg \omega_c$.

%%%%%%%%%%%%%%%%%%%%%%%%%%%%%%%%%%%%%%%%%
\section{Size distribution  of  ATP-islands}
%%%%%%%%%%%%%%%%%%%%%%%%%%%%%%%%%%%%%%%%%

Experiments probing the structure of  single filaments can give 
information not only on the length of the ATP-tip but 
the whole distribution of   ATP-islands sizes in the filament. 
The  average number $I_k$ of ATP-islands 
of length $k$ fulfills   
the master equation 
\begin{eqnarray}
\partial_t I_k &=&J_T(p_{k-1}-p_k) 
- \omega_c \left(1+ (k-1)\rho_c\right)I_k
\nonumber\\
 && +\omega_c(1+ \rho_c) I_{k+1}
    + 2\omega_{c}\rho_c \sum_{s\ge k+2} I_s
\label{Ikcoop}
\end{eqnarray}
for $k \ge 1$. 
The first term on the rhs of eq.\ (\ref{Ikcoop}) 
gives the change in ATP-island numbers 
from attachment and detachment at  the tip.
 The second  term  describes 
 the loss  from cleavage at the $k-1$ 
 sites of the ATP-island with $T$-neighbors on the pointed side 
with rate $\omega_{cT}$ and 
the loss  from cleavage with rate $\omega_{cD^*}$ at 
 the ${\rm TD}^*$-interface at the pointed side of the island. 
The third and fourth terms on the rhs of eq.\ (\ref{Ikcoop})  are 
gain terms from cleavage at an interior site of an island.
The fourth term 
 means that any ATP-island  of length $k$ can 
 be obtained by ``fragmentation'' of an island of length $s\ge k+2$ 
 in two  ways with a rate $\omega_{cT}$ for each way. 
 The third term means that an  ATP-island  of length $k$ can also 
 be obtained  by cleavage at the 
 boundaries of an  ATP-island of length $k+1$ with rate $\omega_{cT}$ on 
the barbed side and with rate $\omega_{cD^*}$ on the pointed side 
of the island.

Two important quantities, which follow from the 
ATP-island distribution and can be observed in experiments on single 
actin filaments, are the average total  number of ATP-islands 
$I \equiv \sum_{k\ge 1} I_k$
and the average total number of ATP-actin protomers
$\langle N_T \rangle \equiv \sum_{k\ge 1} k I_k$.
The average total number $\langle N_T \rangle$ of ATP-actin protomers
gives a measure of the average total length of the ATP-cap of the 
actin filament. The  average total  number 
of ATP-islands
$I$  also gives the number of $\rm TD^*$-interfaces 
within the ATP-cap,  where cleavage takes place with rate 
$\omega_{cD^*}$. 
Knowledge of these two quantities therefore 
not only characterizes the filament structure but allows us to 
also calculate the resulting cleavage flux.
For the two quantities $I$ and $\langle N_T \rangle$, 
we obtain the rate equations 
\begin{align}
\partial_t I &= 
  J_T p_0
       + \omega_{c}\rho_c (\langle N_T\rangle-2I) -
       \omega_c (1-\rho_c)I_1
  \label{I}\\
\partial_t \langle N_T \rangle &=
 J_T -\omega_c\rho_c(\langle N_T\rangle-I)
   - \omega_cI
\label{N}
\end{align}
Equation (\ref{I}) describes the creation of 
 additional ATP-islands 
by addition of T-monomers at an empty tip or by cleavage in the 
interior of an existing T-island with rate $\omega_{cT}$,
whereas  the island number
is reduced by cleavage of ATP-islands of unit length  with a rate 
$\omega_{cD^*}$.
The first term on the rhs of  eq.\  (\ref{N}) describes the  addition of 
  T-monomers with the rate $J_T$, the last two terms 
the loss  of T-protomers by cleavage:
the total number of  ${\rm TD}^*$-interfaces 
with a cleavage rate $\omega_{cD^*}$ is given by $I$, whereas the  
number of sites where cleavage 
happens with the slower rate  $\omega_{cT}$
is given by  $\langle N_T\rangle-I$. 
Therefore, the total   cleavage flux  is given by 
\begin{equation}
J_c = \omega_{c}\rho_c(\langle N_T\rangle-I)  + \omega_{c}I
\label{Jc}
\end{equation}
and eq.\ (\ref{N}) is equivalent to 
 $\partial_t \langle N_T \rangle =  J_T-J_c$:
In a steady state we must have 
$J_T = J_c$, i.e., T-protomer 
addition flux and cleavage flux  balance.

\begin{figure*}
  \begin{center}
  \epsfig{file=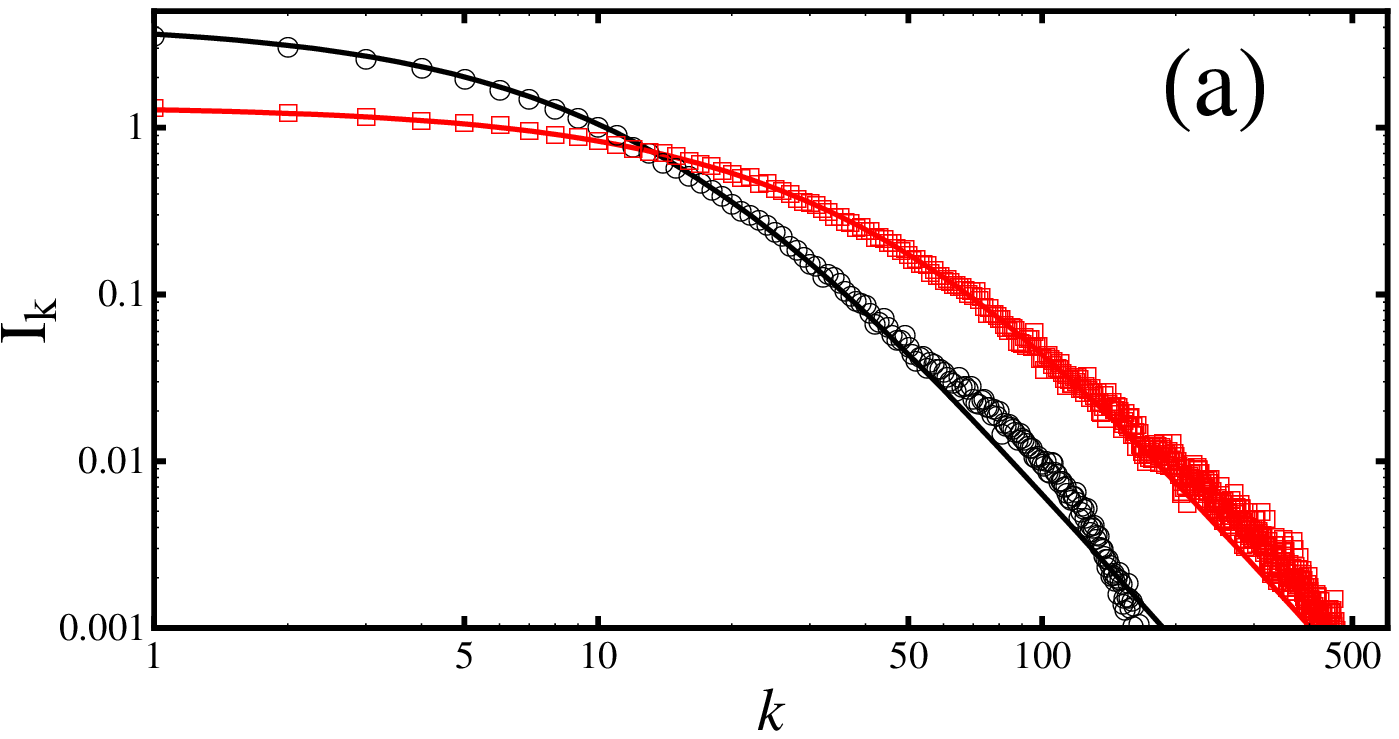,width=0.48\textwidth}~~
  \epsfig{file=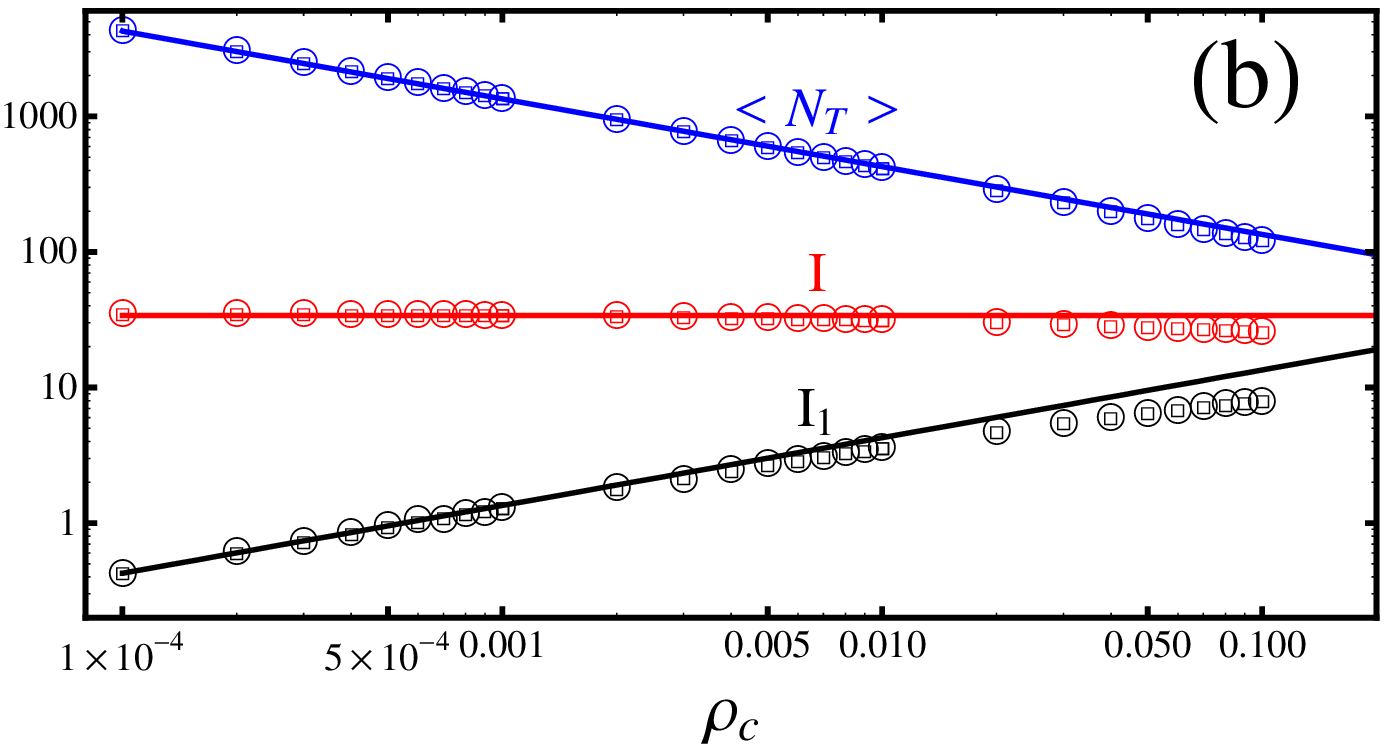,width=0.48\textwidth}
  \caption{
(a) Double-logarithmic ploat of the size distribution 
$I_k$ of ATP-islands   for  an actin concentration 
$C_T=1\mu M$ (corresponding to $J_T/\omega_c = 34$)
and cleavage  parameters $\rho_c=10^{-2}$ (black, $\circ$) and $10^{-3}$ 
(red, $\square$).
Other parameter values are as in table \ref{tableparameters}.
Comparison between (i) analytic results from  eq.\ (\ref{Ikresult3})  
(solid lines) and 
(ii)  results from  stochastic simulations 
 using the Gillespie algorithm (circles).
(b)   Double-logarithmic plot of 
$\langle N_T\rangle$ (blue), 
$I$ (red), and $I_1$ (black) 
 in the steady state  as a function of $\rho_c$ for 
$C_T=1\mu M$ ($J_T/\omega_c = 34$).
Comparison between (i) analytic results from eq.\ (\ref{Ikresult3}) 
 (solid lines), 
 (ii) results from a stochastic simulations
using the Gillespie algorithm (circles), and 
(iii) results from a numerical integration of  the full 
master equations (\ref{nkcoop}) and  (\ref{Ikcoop}) (squares).
    }
  \label{plot2}
  \end{center}
\end{figure*}

The steady state of the ATP-island 
distribution $I_k$ fulfills   $\partial_t I_{k} = 0$.
We determine the stationary island distribution 
for fast growth  in two steps:
(i) We obtain a differential equation for 
$I_k$ in $k$ by requiring that  $\partial_t (I_{k} - I_{k-1})=0$.
This will give the stationary $I_k$ apart from one integration constant. 
(ii) We will determine this integration constant by looking for stationary 
solutions of eq.\ (\ref{N}) for   $\langle N_T \rangle$. 

i) To determine the steady-state  ATP-island distribution
from eq.\ (\ref{Ikcoop}),    
we  consider $\partial_t (I_{k} - I_{k-1})$  for $k\ge 2$
 and use continuous  $k$, which leads to
\begin{align}
\partial_t (I_{k} -I_{k-1} )
 {} \approx &
  - J_T \partial_k^2p_k(k-1)
 +\left( \omega_{c}- \omega_{c}\rho_c \right)\partial_k^2I_k(k) 
\nonumber\\
 &
 -\omega_{c}\rho_c (k-1) \partial_k I_k(k) 
  -3\omega_{c}\rho_cI_k 
.
\end{align}
The steady state fulfills  $\partial_t (I_{k} -I_{k-1})=0$.
In the limit of fast growth, the term 
$J_T \partial_k^2p_k$ is exponentially small, 
and we find 
\begin{equation}
 I_k = c_I  
     {\cal I}\left({\sqrt{2\rho_c}(k-1)}\right).
\label{Ikresult1}
\end{equation}
The integration constant  $c_I$
 has yet to be determined.
The  scaling function 
 ${\cal I}(x) =
       2^{-3/2} D_{-3}(x/\sqrt{2}) e^{x^2/8}$,
where $D_\nu(x)$ is Whittaker's parabolic cylinder function \cite{abra84},
fulfills the differential equation
\begin{equation}
   0= 2{\cal I}''(x) - x {\cal I}'(x) -3{\cal I}(x).
\label{IDGL}
\end{equation}
It decays as ${\cal I}(x) \approx x^{-3}$ for $x\gg 1$,
which  
gives rise to a power-law tail in the island distribution 
with the same scaling behavior 
 $I_k \sim k^{-3}$ as for random cleavage \cite{A07}. 
From the island distribution (\ref{Ikresult1}), we derive
\begin{align}
  I &\approx  \frac{c_I}{\sqrt{2\rho_c}} \int_{0}^\infty dx {\cal I}(x)
 = \frac{c_I}{4\sqrt{2\rho_c}}  
\nonumber\\
 \langle N_T \rangle &\approx  \frac{c_I}{2\rho_c}
    \int_{0}^\infty dx x {\cal I}(x)
    = \frac{\sqrt{\pi}c_I}{8\rho_c} 
\label{Nresult2}
\end{align}
where we used the relations
$\int_0^\infty dx {\cal I}(x) = -{\cal I}'(0)= {1}/{4}$
and $\int_0^\infty dx x{\cal I}(x)   =2{\cal I}(0)=  {\sqrt{\pi}}/{4}$,
which follow from the differential equation   (\ref{IDGL}).

ii) Using the results (\ref{Nresult2}) in eq.\ (\ref{N}) and requiring 
stationarity $\partial_t \langle N_T \rangle =0$ to leading order in $\rho_c$,
 we determine the integration constant 
\begin{equation}
 c_I \approx  4\sqrt{2} {J_T\sqrt{\rho_c}}/{\omega_c}. 
 \label{c_I}
\end{equation}
This leads to  the final results
\begin{align}
  I_1 &\approx   \sqrt{\frac{\pi}{2}}
         \frac{J_T \sqrt{\rho_c} }{\omega_c} 
,~~~
  I \approx  \frac{J_T}{\omega_c} 
,~~~
 \langle N_T \rangle \approx
     \sqrt{\frac{\pi}{2}} \frac{J_T}{\omega_c\sqrt{\rho_c}}.
\label{Ikresult3}
\end{align}
for the average total length $\langle N_T \rangle$ of the ATP-cap, 
the total number of ATP-islands $I$ and the number $I_1$ 
of short ATP-islands of unit length, see Fig.~\ref{plot2}.
The result $I\approx J_T/\omega_c \gg 1$ for the total number
 of ATP-islands  is remarkable 
because we have $I=1$ for strictly vectorial cleavage. Thus a small 
cleavage parameter $\rho_c\neq 0$ 
constitutes a singular perturbation, which 
gives rise to a pronounced change in the ATP-cap structure with a
dramatic increase in the total number of ATP-islands.

%%%%%%%%%%%%%%%%%%%%%%%%%%%%%%%%%%%%%%%%%
\section{ATP-cap length and cleavage flux}
%%%%%%%%%%%%%%%%%%%%%%%%%%%%%%%%%%%%%%%%%

The scaling behavior of $I_k$ in eq.\ (\ref{Ikresult3}) 
leads to a characteristic scaling of island sizes $k$ with the 
square root of the cleavage parameter $\rho_c$,
which determines the $\rho_c$-dependence of 
experimental observables such as the average total length of the ATP-cap 
length $\langle N_T \rangle$ and the cleavage flux $J_c$.  

\begin{figure*}
  \begin{center}
  \epsfig{file=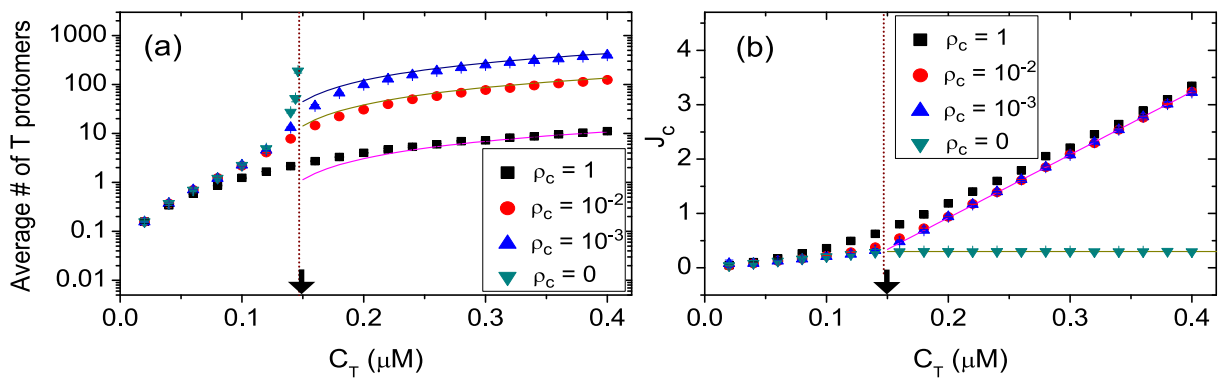,width=0.9\textwidth}
 \caption{
(a) The total average ATP-cap length $\langle N_T\rangle$ and 
(b) the total cleavage flux $J_c$ as a function of actin 
concentration $C_T$ for different cleavage parameters $\rho_c$. 
Comparison between (i) analytic results for $C_T\gg C_{T,c}$ 
(the critical concentration $C_{T,c}$ is marked by an arrow) in the 
regime of fast growth (solid lines) and 
and (ii) results from a stochastic simulations 
using the Gillespie algorithm (data points).
    }
  \label{capeps}
  \end{center}
\end{figure*}

According to (\ref{Ikresult3})  the ATP-cap length $\langle N_T \rangle$
 depends linearly on the T-protomer growth rate $J_T$ and, thus, 
on the  concentration $C_T$ of T-monomers and increases $\propto
1/\sqrt{\rho_c}$ 
for small cleavage parameters corresponding to 
 strongly cooperative
 cleavage mechanisms, see Fig.\ \ref{capeps}(a). 
This result is corroborated by a scaling argument \cite{XLJKRL}.  
It shows that
experiments on the total cap length in the limit of fast growth 
will allow to determine the cleavage parameter $\rho_c$
 if the cleavage rate is known. 
On the other hand,  
the cap length in (\ref{Ikresult3}) only depends on the {\em product} 
$1/\omega_c\sqrt{\rho_c}=1/\sqrt{\omega_{c,D^*}\omega_{c,T}}$. 
Thus, a mechanism with high cooperativity and large cleavage rate 
can give rise to a similar cap length as a random cleavage 
process with a low cleavage rate. This demonstrates that  
measurements of certain  
 filament properties such as the ATP-monomer content 
of filaments  do not allow to uniquely 
distinguish between  a  
 vectorial model with high cleavage 
rate \cite{PCK85,korn87,melki96,SK06} and a 
random model with lower cleavage rate \cite{blanchoin02,VYOS05,fuji07}.

 From the coupled 
rate eqs.\  (\ref{I}) and (\ref{N}) for $I$ and $\langle N_T \rangle$,
we can also calculate 
the characteristic time scale $\tau \approx  1/\omega_c\sqrt{\rho_c}$
for relaxation to the steady state with 
stationary values  (\ref{Ikresult3}). 
This experimentally relevant time scale is  increasing 
with the same $\rho_c^{-1/2}$-dependence 
as the steady state ATP-cap length for small cleavage parameters.

According to eqs.\ (\ref{N}) and 
(\ref{Jc}),  the cleavage flux  $J_c$ and  the T-protomer 
addition flux $J_T$  must balance in the  steady state, i.e., 
$J_c =  J_T$.
Because of   eq.\ (\ref{JT})
the cleavage flux depends then  {\em linearly} on 
the T-monomer concentration $C_T$  and becomes 
{\em independent} of the cooperativity parameter
for fast growth with $P_{1,T}\approx 1$, see Fig.\ \ref{capeps}(b).
 For vectorial cleavage, 
on the other hand, we have $I=1$,
 and the cleavage flux is directly given by 
the cleavage rate, $J_c = \omega_c$ and, thus, {\em independent}
of the T-monomer concentration, see Fig.\ \ref{capeps}(b). 
This shows again that a non-zero cleavage parameter 
represents a singular perturbation of vectorial cleavage, 
and the cleavage flux is a sensitive quantity to differentiate 
between strictly vectorial cleavage  with 
$\rho_c=0$  and strongly cooperative 
cleavage with small but nonzero $\rho_c$.

%%%%%%%%%%%%%%%%%%%%%%%%%%%%%%%
\section{Conclusion}

In conclusion, we have introduced  an effective two-state model 
for cooperative ATP-hydrolysis in actin filaments, 
where cooperativity is characterized 
by the cleavage parameter $\rho_c$. The model  contains random 
 ($\rho_c=1$) and vectorial ($\rho_c=0$) cleavage as special cases. 
For this two-state model, 
 we could obtain 
analytic steady-state results for quantities such as the 
size of the ATP-actin cap,  the size distribution of ATP-actin 
islands, the total number of ATP-actin islands, which describe the 
structure of the actin filament, and kinetic quantities such as 
the cleavage flux.
Measurements of these steady state quantities will allow 
to determine the cleavage rate and the cooperativity
of the ATP-cleavage mechanism. 
Recently depolymerization experiments on individual actin filaments 
have become possible \cite{fuji07,kueh08}, which allow 
to obtain information on filament structure based on the different
depolymerization rates of T-, $\Theta$- and D- monomers. 
The depolymerization experiments in Ref.\ \cite{kueh08}
as analyzed in Ref.\ \cite{XLJKRL} and earlier kinetic data 
discussed in Refs.\  \cite{carlier87,korn87} suggest that the cleavage 
parameter $\rho_c$ is as small 
as $10^{-5}-10^{-6}$ and  that cleavage is thus strongly cooperative. 
Our analytic results become exact in the limit of small $\rho_c$ 
and fast growth $J_T\gg \omega_c$
and, thus, can be directly applied in this relevant  parameter regime. 
Measurements of force-velocity relations 
 for polymerization under force can  also provide sensitive probes 
of the filament structure as discussed in Ref.\ \cite{ranjith09}.

%%%%%%%%%%%%%%%%%%%%%%%%%%%%%%%%%%

\end{document}